\documentclass[prd,reprint,nofootinbib,showpacs,superscriptaddress]{revtex4-2}
\usepackage{graphicx} 
\usepackage{hyperref}
\usepackage{amsfonts}
\usepackage{amsmath,amssymb}
\usepackage{bm} 
\usepackage{color}
\usepackage{epstopdf} 
\usepackage{epsfig}
\usepackage{subfig} 
\usepackage{float}

\usepackage[justification=raggedright,singlelinecheck=false]{caption}

{\rm }

\def\be{\begin{equation}}
 \def\ee{\end{equation}}
 \def\bea{\begin{eqnarray}}
 \def\eea{\end{eqnarray}}
 \def\bes{\begin{eqnarray}}
 \def\ees{\end{eqnarray}}
 \def\bi{\begin{itemize}}
 \def\ei{\end{itemize}} 

\def\2{\frac{1}{2}}
\def\4{\frac{1}{4}}

\begin{document}

\title{Detector noise in continuous-variable quantum key distribution}

\author{Shihong Pan}
\email{sp5560@nyu.edu}
\affiliation{New York University Shanghai, 567 West Yangsi Road, Shanghai, 200126, China}

\author{Dimitri Monokandylos}
\affiliation{New York University Shanghai, 567 West Yangsi Road, Shanghai, 200126, China}
\affiliation{The University of Edinburgh, 57 George Square Edinburgh EH8 9JU, Scotland}

\author{Bing Qi}
\email{bing.qi@nyu.edu}
\affiliation{New York University Shanghai, 567 West Yangsi Road, Shanghai, 200126, China}
\affiliation{NYU-ECNU Institute of Physics at NYU Shanghai, 567 West Yangsi Road, Shanghai, 200126, China}
\affiliation{State Key Laboratory of Precision Spectroscopy, School of Physical and Material Sciences, East China Normal University, Shanghai, 200062, China}

\date{\today}

\begin{abstract}
{In continuous-variable (CV) QKD with optical coherent detection, the widely adopted \textit{trusted detector noise} model improves both the secret key rate and the transmission distance. This model assumes that detector noise is inherently random and inaccessible to an adversary. While substantial research has focused on shielding the detector, it is far more difficult to justify the adversary's ignorance of the detector noise. In this paper, we introduce a \textit{calibrated detector noise} model for CV-QKD, which relies solely on the isolation of the detector from the adversary's intervention. Specifically, our model applies even when detector noise is predictable to the adversary. We analyze the electrical noise of a commercial balanced photoreceiver and perform numerical simulations to compare different noise models. Our results show that when the detector noise variance is an order of magnitude below the vacuum noise, the proposed model achieves a secret key rate comparable to that of the trusted detector noise model, while eliminating the questionable assumption of ``truly random'' detector noise.}

\end{abstract}

\maketitle

\section{Introduction}
\label{sec:1}

One important application of quantum information science is quantum key distribution (QKD), a protocol that enables two remote parties (Alice and Bob) to establish cryptographic keys with proven security \cite{gisin2002quantum, diamanti2016practical, xu2020secure, pirandola2020advances}. 

The security of QKD is rooted in the fundamental laws of quantum mechanics. When a QKD protocol is properly designed and implemented, any eavesdropping attempt by an adversary (Eve) that yields a non-negligible amount of information will inevitably introduce noise, thereby revealing her presence. The security proof of a given QKD protocol enables the legitimate users to quantify Eve’s potential information gain based on the observed noise and other system parameters, allowing them to estimate the length of the final secret key. In general, higher observed noise indicates greater potential information leakage, leading to a shorter secret key.

In practice, no QKD implementation is perfect. Real-world imperfections introduce system noise even in the absence of Eve. If the legitimate users cannot distinguish intrinsic system noise from noise caused by Eve, a conservative approach is to attribute all observed noise to Eve’s attack and estimate her potential information gain accordingly. While this ``untrusted noise'' model may overestimate Eve's information and reduce QKD performance, it minimizes reliance on additional assumptions about the system. Consequently, it is commonly used in discrete-variable (DV) QKD based on single-photon detection \cite {bennett2014quantum}, where the system noise is at a low level.

In principle, the untrusted noise model described above can also be applied to continuous-variable (CV) QKD based on coherent detection, a potentially cost-effective solution compatible with classical fiber-optic networks \cite{ralph1999continuous, hillery2000quantum, grosshans2003quantum}. However, this approach often leads to poor performance, particularly under high channel loss. The primary limiting factor is the electrical noise of the optical coherent detector, which is typically an order of magnitude lower than vacuum noise \cite{Chi2011, huang2013300, kumar2012versatile}. In such cases, attributing detector noise to Eve's attack significantly reduces both the secret key rate and the achievable QKD distance. 

One approach to mitigate this issue is to adopt the trusted detector noise model, which is widely used in practical CV-QKD implementations \cite{grosshans2003quantum, lodewyck2007quantum, qi2007experimental, jouguet2013experimental, huang2016long, zhang2019continuous, hajomer2025chipbasedcvqkd, wang2025high}. In this model, detector noise is treated as an intrinsic property that cannot be controlled or accessed by Eve. By distinguishing such intrinsic detector noise from other untrusted noise—potentially introduced by Eve—and treating them differently in the security analysis, a tighter bound on information leakage can be obtained, leading to improved QKD performance. The trade-off, however, is that the legitimate users must make additional assumptions about the QKD system.

The validity of the trusted detector noise model relies on two key assumptions: (1) the detector can be accurately calibrated by the legitimate user and remains isolated from the adversary, and (2) the detector noise is truly random from the adversary's perspective. While substantial research has focused on strengthening the first assumption \cite{jouguet2013preventing, zhang2020one, brunner2020precise, zou2022rigorous}, the second is far more difficult to justify. In principle, only quantum noise can be considered truly random and thus suitable for cryptographic applications. However, verifying the quantum origin of the electrical noise in a real-life optical coherent detector remains challenging. In fact, in quantum random number generation (QRNG) using optical homodyne detection, detector noise is often treated as ``untrusted'' and unsuitable for generating true randomness \cite{gabriel2010generator, qi2010high, ma2013postprocessing, zheng20196, bai202118, zhang2024selftestingqrng,wang2025highly}.

In this paper, we introduce a \textit{calibrated detector noise} model for CV-QKD, which relies solely on the first assumption. Specifically, our model applies even when detector noise is predictable to the adversary. We analyze the electrical noise of a commercial balanced photoreceiver and perform numerical simulations to estimate the asymptotic secret key rates of the Gaussian-modulated coherent states (GMCS) QKD protocol \cite{grosshans2003quantum} under different noise models. Our results show that the proposed model can achieve a secret key rate comparable to that of the trusted-noise model, while removing the questionable assumption of ``truly random'' detector noise.

This paper is organized as follows: Section \ref{sec:2} reviews two existing noise models in CV-QKD—the trusted noise model and the untrusted noise model. In Section \ref{sec:3}, we experimentally evaluate the electrical noise of a commercial balanced photoreceiver. Section \ref{sec:4} introduces a new noise model for CV-QKD: the calibrated detector noise model. In Section \ref{sec:5}, we present numerical simulation results of secret key rates under different noise models. Finally, Section \ref{sec:6} concludes the paper with a discussion.

\section{Trusted and untrusted detector noise models in CV-QKD}
\label{sec:2}

The basic procedures of the GMCS QKD without switching \cite{weedbrook2004quantum} are as follows: for each transmission, Alice draws two random numbers $x_A$ and $p_A$ and transmits the coherent state $|x_A+ip_A\rangle$ to Bob through an insecure quantum channel. Here $x_A$ and $p_A$ are Gaussian random numbers with zero mean and a variance of $V_A N_0$, where $V_A$ is the modulation variance and $N_0$ = 1/4 denotes the shot-noise variance. Upon receiving the state, Bob performs a simultaneous measurement of both quadratures X and P using a conjugate homodyne detection setup. By repeating this quantum communication process multiple times, Alice and Bob collect sufficient data to estimate the channel transmittance and the level of excess noise via an authenticated classical channel. If the observed noise is below a predetermined security threshold, they proceed with key extraction by applying error reconciliation followed by privacy amplification, ultimately generating a shared secret key.

In general, the performance of CV-QKD is determined by both the transmittance and noise characteristics of the quantum channel, as well as the efficiency and intrinsic noise of the QKD system itself. It is standard practice to distinguish between two types of noise: channel noise, which may result from Eve’s intervention and is characterized by a variance denoted as $\Xi_\text{ch}$, and detection noise, which originates from imperfections in the QKD receiver and is inaccessible to Eve, with variance of $\Xi_\text{det}$. Following common conventions, $\Xi_\text{ch}$ refers to the channel input while $\Xi_\text{det}$ refers to Bob's input. All noise quantities are expressed in shot-noise units (SNU).

Regardless of the specific noise model, the total noise variance $\Xi_\text{tot}$, refers to the channel input, can be expressed as
\begin{equation}
    \Xi_\text{tot} = \Xi_\text{ch} + \frac{\Xi_\text{det}}{T}
    \label{eq:totalNoise}
\end{equation}
where $T$ is the transmittance of the channel.

Given the above parameters, the secret key rate of the GMCS QKD protocol can be determined. In this paper, we consider the protocol using reverse reconciliation and conjugate homodyne detection scheme. The general formula for the lower bound of the asymptotic key rate is then given by \cite{Devetak2005,lodewyck2007quantum}
\begin{equation}
    R=fI_{AB}-\chi_{BE},
    \label{eq:keyRate}
\end{equation}
where $f$ is the reconciliation efficiency; $I_{AB}$ is the mutual information between Alice and Bob; and $\chi_{BE}$ is the Holevo quantity which upper bounds the information between Eve and Bob.

The mutual information $I_{AB}$ depends only on the total noise variance $\Xi_\text{tot}$, and not on how it is distributed between $\Xi_\text{ch}$ and $\Xi_\text{det}$. Therefore, regardless of the noise model adopted, the mutual information between Alice and Bob is given by
\begin{equation}
    I_{AB}=\log_2(\frac{V+\Xi_\text{tot}}{1+\Xi_\text{tot}}),
    \label{eq:mutualInfo}
\end{equation}
where $V=V_A+1$. In the above formula, we assume that both quadratures, $X$ and $P$, are used to generate the secret key.

In general, calculating the Holevo bound $\chi_{BE}$ is nontrivial and depends on the specific noise model adopted. Under collective attacks, the Holevo bound for GMCS QKD is given by \cite{lodewyck2007quantum,Fossier2009}
\begin{equation}
    \chi_{BE}=\sum_{i=1}^2G(\frac{\lambda_i-1}{2})-\sum_{i=3}^5G(\frac{\lambda_i-1}{2})
    \label{eq:holevoBound}
\end{equation}
where $G(x)=(x+1)\log_2(x+1)-x\log_2x$. The values of $\lambda_i$ are given by \cite{lodewyck2007quantum,Fossier2009}
\begin{equation}
    \begin{aligned}
        \lambda_{1,2}^2&=\frac{1}{2}(A\pm\sqrt{A^2-4B})\\
        \lambda_{3,4}^2&=\frac{1}{2}(C\pm\sqrt{C^2-4D})\\
        \lambda_5&=1
        \label{eq:symplecticValues}
    \end{aligned}
\end{equation}
where
\begin{equation}
    \begin{aligned}
        A&=V^2(1-2T)+2T+T^2(V+\Xi_\text{ch})^2\\
        B&=T^2(V\Xi_\text{ch}+1)^2\\
        C&=(\frac{1}{T(V+\Xi_\text{tot})})^2[A\Xi_\text{det}^2+B+1+\\&2\Xi_\text{det}(V\sqrt{B}+T(V+\Xi_\text{ch}))+2T(V^2-1)]\\
        D&=(\frac{V+\sqrt{B}\Xi_\text{det}}{T(V+\Xi_\text{tot})})^2.
        \label{eq:ABCD}
    \end{aligned}
\end{equation}
With these quantities, the secret key rate $R$ can be calculated.

Note in Eq. \ref{eq:ABCD} the channel noise and the detection noise depend on the noise model employed. In the trusted detector noise model, Eve has no control over Bob's setup, and the noises of Bob's detectors are assumed to be truly random to Eve. The channel noise $\Xi_\text{ch}^\text{trusted}$ therefore only depends on the channel transmittance $T$ and noises outside Bob's detector, and is given by \cite{Scarani2009}
\begin{equation}
    \Xi_\text{ch}^\text{trusted} = \frac{1-T}{T} + \xi_A\\
    \label{eq:channelNoiseTrusted}
\end{equation}
where $\xi_A$ represents both the excess noise contributed by Alice during her state preparation and the excess noise from the channel. Similarly, the detection noise $\Xi_\text{det}^\text{trusted}$ on Bob's side relies on the detector efficiency and the electrical noise of the detector. Assuming conjugate homodyne detection scheme, the detection noise is given by
\cite{Fossier2009}
\begin{equation}
    \Xi_\text{det}^\text{trusted} = \frac{1+(1-\eta_B)+2\nu_B}{\eta_B}
    \label{eq:detectionNoiseTrusted}
\end{equation}
where $\eta_B$ is Bob's detector efficiency and $\nu_B$ is the electrical noise of the detector. 

In the untrusted detector noise model, it is assumed that Eve can manipulate Bob’s detection noise, though not the intrinsic loss of Bob’s detector. As a result, the electrical noise of Bob’s detector can be included into the channel noise, leading to modified noise parameters: channel noise $\Xi_\text{ch}^\text{untrusted}$ and detection noise $\Xi_\text{det}^\text{untrusted}$:
\begin{equation}
    \begin{aligned}
        \Xi_\text{ch}^\text{untrusted} &= \frac{1-T}{T} + \xi_A + \frac{2\nu_B}{T\eta_B}\\
        \Xi_\text{det}^\text{untrusted} &= \frac{1+(1-\eta_B)}{\eta_B}.
    \label{eq:noiseUntrusted}
    \end{aligned}
\end{equation}
The key rates for the two models are obtained by substituting the respective noise values into Eqs. \ref{eq:mutualInfo} and \ref{eq:holevoBound}.

\section{Experimental characterization of optical homodyne detector}
\label{sec:3}

Optical homodyne detectors, widely used in CV quantum communication protocols, are primarily characterized by their detection efficiency, bandwidth, and electrical noise. While some of the electrical noise may originate from quantum sources, it is often challenging in practice to precisely identify and separate its exact origins.

\begin{figure}[H]
    \includegraphics[width=0.5\textwidth]{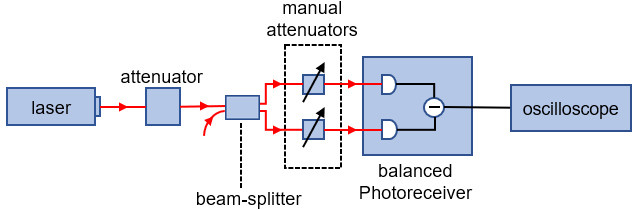}
    \caption{Experimental setup. The local oscillator is attenuated to a desired power before entering a 50/50 beam-splitter with a vacuum state. The outputs then pass through manual attenuators before being measured by a balanced photoreceiver. The differential signal from the photoreceiver is sampled using a 200 MHz digital oscilloscope.}
    \label{fig:homodyne}
\end{figure}

In this section, we characterize the performance of a well-designed commercial balanced detector. More specifically, we operate the detector under varying local oscillator (LO) power levels and examine the resulting output of the homodyne detector. The experimental setup is illustrated in FIG. \ref{fig:homodyne}. The output of a 1550 nm narrow-linewidth laser is passed through a programmable optical attenuator and then fed into one input of a 50/50 fiber beam splitter, serving as the LO for homodyne detection. The other input of the beam splitter receives a vacuum state. The two outputs of the beam splitter are carefully balanced using manual optical attenuators and then directed to a balanced photoreceiver (FEMTO, model HBPR-500M-10K-IN-FC). The programmable attenuator allows precise control of the LO power. The differential signal from the balanced photoreceiver is sampled using a 12-bit, 200 MHz digital oscilloscope. 

\begin{figure}
    \hspace*{-0.7cm}
    \includegraphics[width=0.5\textwidth]{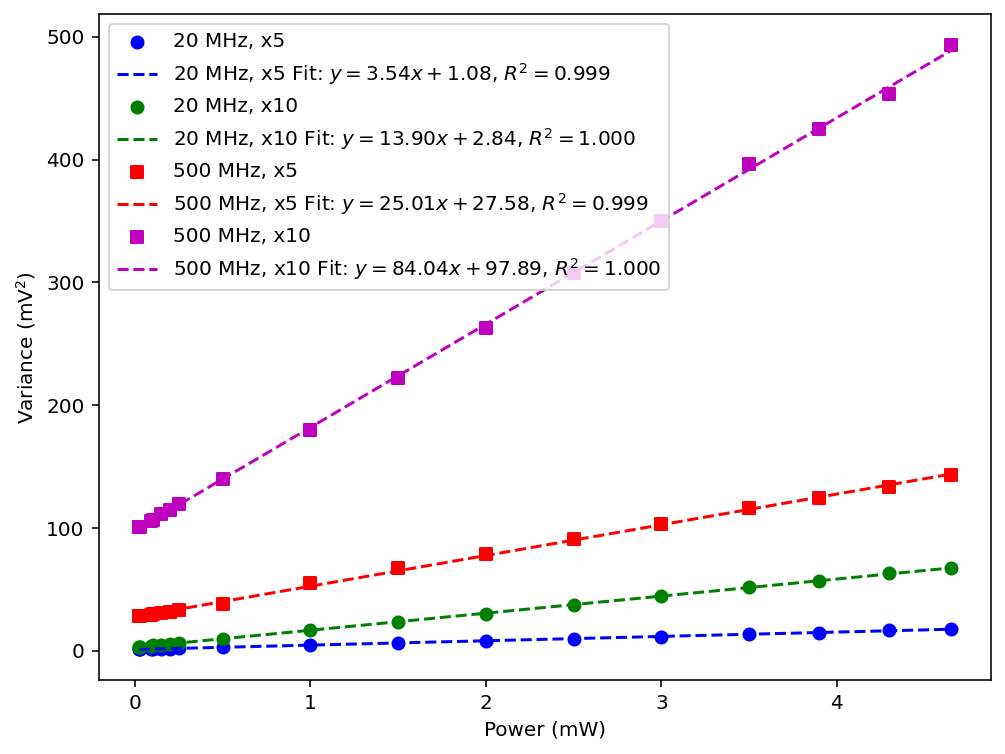} 
    \centering
    \caption{Homodyne detection of a vacuum state at different local oscillator powers, with a detection bandwidth of 20 MHz (circular points) at gains of $5\times10^3$ V/A (blue) and $10\times10^3$ V/A (green), and a bandwidth of 500 MHz (square points) at gains of $5\times10^3$ V/A (red) and $10\times10^3$ V/A (purple). The dashed lines represent linear fits to the experimental data, with the corresponding fitting equations shown in the legend.}
    \label{fig:variance_vs_power}
\end{figure}

The differential output of the balanced photoreceiver represents the quadrature value of the vacuum state, from which perfect randomness could be generated \cite{gabriel2010generator}. Shown in FIG.\ref{fig:variance_vs_power} is the total variance of the detector output, $\sigma_T^2$, that was measured at different LO powers, for different detector setups (bandwidth of 20 MHz or 500 MHz, gain of $5\times10^3$ V/A or $10\times10^3$ V/A). 

 In general, the total noise observed is contributed by both the vacuum noise and the electrical noise of the homodyne detector, with corresponding noise variance of $\sigma^2_{Q}$ and $\sigma^2_{E}$. Note 
 $\sigma_E^2$ can be measured when the LO power is zero, as shown in FIG. \ref{fig:variance_vs_power}. The detector noise $\nu_B$ defined in Eq. \ref{eq:detectionNoiseTrusted} is $\sigma^2_{E}$ expressed in the shot noise unit.
 
 Assuming that the vacuum noise is independent of the electrical noise of the detector \cite{Feihu2012}, the relationship among the three noise variances can be expressed as:
\begin{equation}
    \sigma^2_{T} = \sigma^2_{E}+ \sigma^2_{Q}
    \label{eq:totalvariance}
\end{equation}

In QRNGs using optical homodyne detection, the vacuum noise is typically treated as quantum noise, while the detector’s electrical noise is assumed to be classical noise. To quantify their relative contributions, the quantum-to-classical noise ratio (QCNR) is defined as \cite{Feihu2012}: 
\begin{equation}
    \mathrm{QCNR} = 10\log_{10}\left(\frac{\sigma_{Q}^2}{\sigma_{E}^2}\right) = 10\log_{10}\left(\frac{\sigma_{T}^2-\sigma^2_{E}}{\sigma_{E}^2}\right)
    \label{eq:qcnr}
\end{equation}
In quantum communication, it is preferable to operate the detector in a high-QCNR regime. 

\begin{figure}[ht]
        \hspace*{-0.7cm}
        \includegraphics[width=0.5\textwidth, height=0.250\textheight]{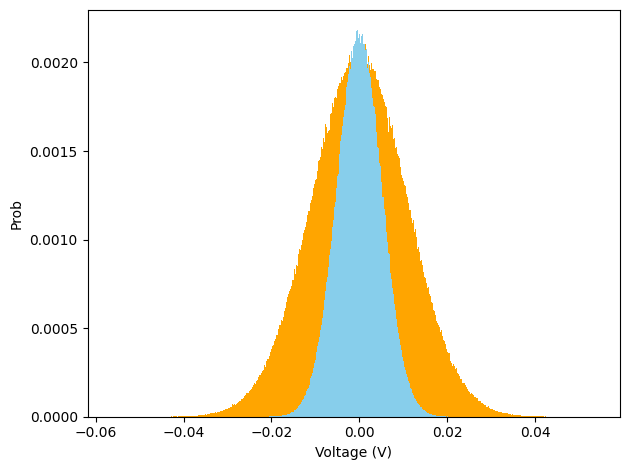}
        \caption{Probability distributions of 5 million homodyne-detected samples acquired at $\mathrm{QCNR} = -\infty$ (blue) and $\mathrm{QCNR} = 5.5$ (orange). Small DC biases have been removed from the samples.}
        \label{fig:pdfhist}
    \end{figure}

The probability distributions of the detector outputs, $x_i$, obtained at a minimal QCNR of $-\infty$ (0 mW LO power), and at a QCNR of 5.5 (using 4.20 mW LO power), are approximately Gaussian, as shown in FIG. \ref{fig:pdfhist}. In these measurements, the detector bandwidth was set to 500 MHz and the sampling rate of the 200MHz oscilloscope was set to 125 MHz.

\begin{figure}[ht]
        \hspace*{-0.7cm}
        \includegraphics[width=0.5\textwidth, height=0.25\textheight]{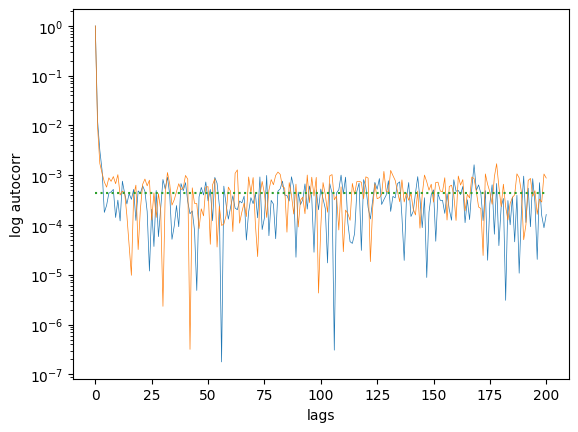}
        \caption{Absolute values of autocorrelation (in log scale) of the homodyne output with $\mathrm{QCNR} = -\infty$ (blue) and $\mathrm{QCNR} = 5.5$ (orange). The dashed green line indicates the expected autocorrelation from a finite set of 5 million truly random numbers, at $\frac{1}{\sqrt{5,000,000}}\approx0.0005$.}
        \label{fig:autocorr}
    \end{figure}

We further calculated the autocorrelations of the samples shown in FIG. \ref{fig:pdfhist}, and the results are presented in FIG. \ref{fig:autocorr}. Notably, the autocorrelations of the total noise (comprising both vacuum noise and detector noise) and of the detector noise alone are at the same level, and their magnitudes match what would be expected from random noise with a finite data size.

It is also worth noting that our experiment presents the basic operation of Bob's detector calibration process. In practice, to deal with unstable detector noises, the calibration could be done in real time by randomly blocking a chunk of signals during the QKD process.

As a brief summary, the experimental results presented in this section suggest that the electrical noise of the detector under test behaves like Gaussian random noise, at least from the legitimate users' perspective. However, in practice, it is difficult to identify the sources of detector noise, making it challenging to justify that the noise is truly random from an adversary's perspective, as assumed in the trusted detector noise model. Could a highly sophisticated adversary gain a deeper understanding of the detector's properties, thereby perceiving it as less noisy? We address this concern by proposing a calibrated detector noise model in the next section.
  
\section{Calibrated detector noise model for CV-QKD}
\label{sec:4}

As discussed earlier, in practice, it is difficult to justify the true randomness of detector noise, a key assumption in the trusted detector noise model. On the other hand, the untrusted detector noise model seems overly pessimistic, as it ignores the fact that the detection system resides within Bob's private space, where practical countermeasures against Eve's interference can be implemented.

Here, we propose a calibrated detector noise model that lies between the untrusted and trusted noise models. On one hand, as in the trusted detector noise model, the detector is assumed to be isolated from external access, and its properties cannot be manipulated by the adversary. On the other hand, the detector noise may not be truly random from Eve’s perspective and, as a conservative assumption, is considered fully predictable to her. A key advantage of this approach is that it removes the need to validate the true randomness of detector noise, a task that is often challenging in practice.

Note in different noise models, we make different assumptions on Eve's capabilities, but the mutual information between Alice and Bob $I_{AB}$ remains unchanged, and thus Eq. \ref{eq:mutualInfo} holds for all the three models. The key difference lies in how detector noise affects Eve's knowledge of Bob's measurement outcomes, i.e. how to calculate $\chi_{BE}$.

In the trusted detector noise model, the detector noise is assumed to be truly random from Eve's perspective. As the detector noise increases, both $I_{AB}$ and $\chi_{BE}$ decrease. Consequently, the secret key rate, which relies on the difference between $I_{AB}$ and $\chi_{BE}$, is only weakly dependent on detector noise, making this model the least sensitive to such noise.

In the untrusted detector noise model, it is assumed that Eve can suppress all intrinsic detector noise, and that all observed excess noise is attributed to her attack. Under this model, as the detector noise increases, the estimated information leakage $\chi_{BE}$ also increases, making this model the most sensitive to detector noise.

Finally, in the calibrated detector noise model where Eve has full knowledge but no control over the detector noise, Bob’s measurement process can be modeled as follows: a noiseless detector performs homodyne detection, after which Gaussian-distributed numbers are classically added to the detector output. The variance of these Gaussian numbers matches the noise variance of the real detector, and the values of the Gaussian numbers are assumed to be known to Eve but unknown to Alice and Bob. As a result, the detector noise affects only $I_{AB}$ but not $\chi_{BE}$.

Based on the above discussion, to calculate Eve’s information $\chi_{BE}$ under the calibrated noise model, the formula presented in Section \ref{sec:2} remains applicable, except that the noise terms must be redefined. Specifically, the channel noise, which represents noise outside Bob's detection system, remains the same as in the trusted model, i.e., $\Xi_\text{ch}=\Xi_\text{ch}^\text{trusted}$, and can be determined using Eq. \ref{eq:channelNoiseTrusted}. However, the detector noise $\Xi_\text{det}$ and the total noise $\Xi_\text{tot}$ from Eve's perspective, which are used in Eq. \ref{eq:ABCD} to calculate Eve's information, should be replaced by 
\begin{equation}
    \begin{aligned}
        \Xi_\text{det}^\text{Eve} &= \frac{1+(1-\eta_B)}{\eta_B}
        \label{eq:noiseRealistic}
    \end{aligned}
\end{equation}
and
\begin{equation}
    \begin{aligned}
        \Xi_\text{tot}^\text{Eve} &= \Xi_\text{ch} + \frac{\Xi_\text{det}^\text{Eve}}{T}.
    \label{eq:totalNoiseRealistic}
    \end{aligned}
\end{equation}

Compared with Eq. \ref{eq:detectionNoiseTrusted}, the term $\nu_B$, which quantifies the detector noise, disappears in Eq. \ref{eq:noiseRealistic}. This indicates that, in the new noise model, Eve’s information about Bob’s measurement results is independent of detector noise. 

\section{Numerical simulation of secret key rates}
\label{sec:5}
\begin{figure}[H]
    \hspace*{-0.7cm}
    \includegraphics[width=0.55\textwidth, height=0.25\textheight]{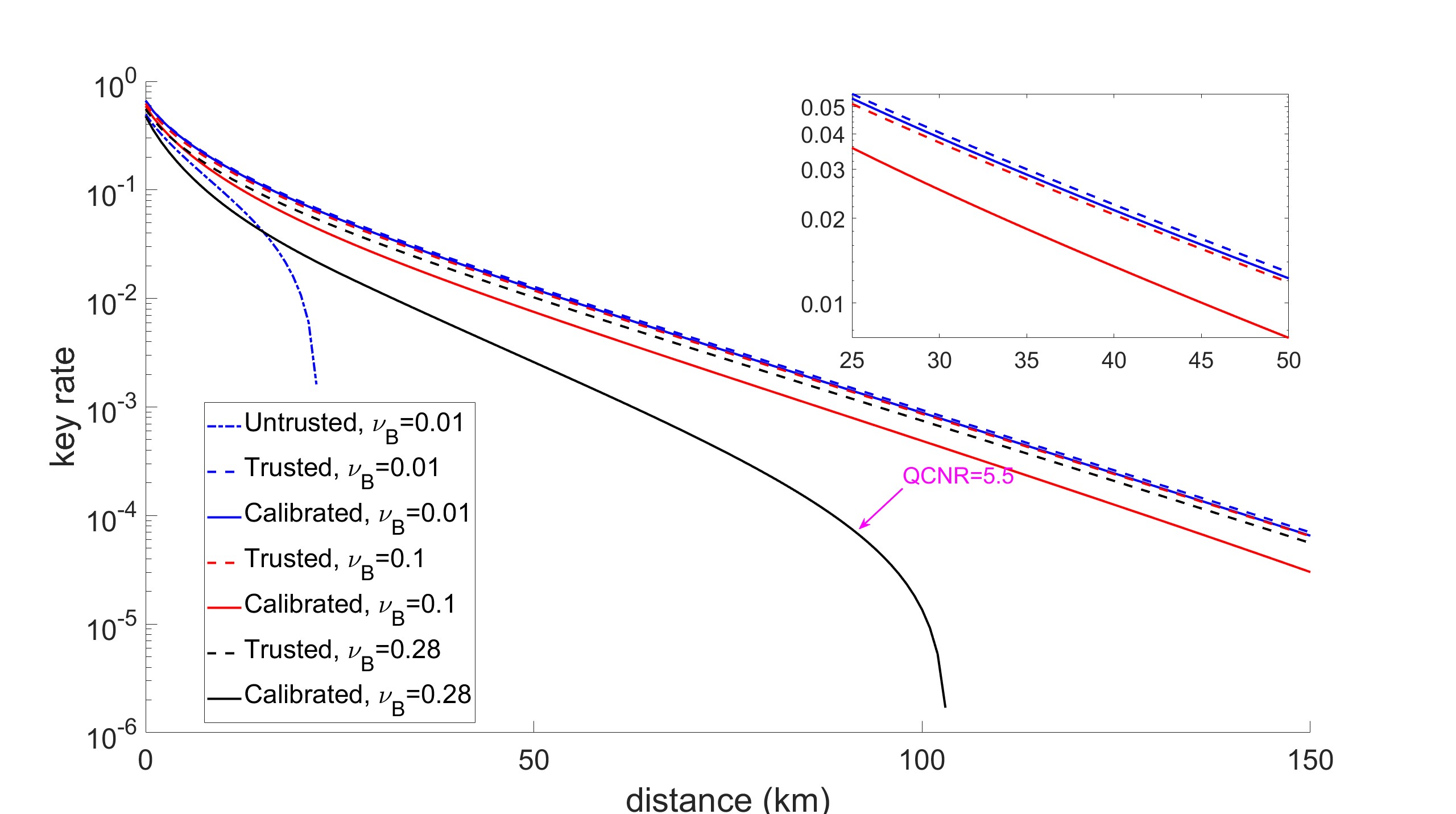}
    \caption{Key rates for GMCS QKD. The three different detector noise models are simulated: trusted (dashed line), untrusted (dash-dotted line), and calibrated (solid line). The electrical noises of the detector are $\nu_B=0.01$ (blue), $\nu_B=0.1$ (red), and $\nu_B=0.28$ (black). The zoom-in subplot shows the key rates in the low noise regime $\nu_B=0.01,0.1$, within the typical length of access networks (25km-50km). The key rates are optimized over the modulation variance $V_A$. Other simulation parameters are: $f=0.95$, $\eta_B=0.5$. The fiber attenuation used is $0.2$dB/km.}
    \label{fig:keyrate_allnoises}
\end{figure}

We perform numerical simulations of the key rates for the GMCS QKD protocol under the three detector noise models, with the reconciliation efficiency $f=0.95$ and the detector efficiency $\eta_B=0.5$. We plot the key rates over 150km with fiber attenuation $0.2$dB/km. For each fiber length, we optimize the key rate over the modulation variance $V_A$. To account for phase noise in a coherent communication system, the excess noise is modeled as $\xi_A=0.01+0.01V_A$ \cite{qi2015generating}. FIG. \ref{fig:keyrate_allnoises} plots the key rates for different electrical noises of the detector: $\nu_B=0.01,0.1,0.28$. 

The case of $\nu_B=0.01$ corresponds to a homodyne detector with very low noise. Achieving such low noise typically requires a detector with a narrow bandwidth \cite{tian2022experimental}. As shown in FIG. \ref{fig:keyrate_allnoises}, although our calibrated detector noise model is based on more realistic assumptions than the trusted detector noise model, the resulting loss in key rate is minimal. Moreover, both the trusted and calibrated models yield significantly higher key rates compared to the untrusted noise model.
The case of $\nu_B=0.1$ represents more realistic detector noise in high-speed CV-QKD systems \cite{Chi2011, huang2013300, kumar2012versatile} while $\nu_B=0.28$ corresponds to the detector described in Section \ref{sec:3} when operated at 500MHz ($\text{QCNR}=5.5$). In these regimes, the key rates for the untrusted detector noise model completely vanish, and the difference between our calibrated noise model and the trusted noise model becomes more visible. $\nu_B=0.1$ yields similar key rates at short distances and deviates slightly at long distances, while $\nu_B=0.28$ shows a significant drop in key rates after 100 km. These simulation results indicate that the calibrated noise model achieves a well-balanced trade-off between key rate and security.

\section{Discussion}
\label{sec:6}
The security of QKD is grounded in both the fundamental laws of quantum physics and a set of assumptions that define the mathematical model of a QKD system. If any of these assumptions are violated, the security proof becomes invalid. It is therefore the responsibility of QKD users to ensure that all assumptions underlying the security proof hold in practice.

In CV-QKD, detector noise is a critical factor affecting system performance. The most conservative approach assumes that Eve can effectively render Bob’s detector noiseless, attributing all observed noise to her attack. This untrusted detector noise model removes the need for detector noise calibration but results in overly pessimistic key rates and is rarely adopted in long-distance CV-QKD.
At the other end of the spectrum, the widely used trusted detector noise model enables higher secret key rates and longer transmission distances. It relies on two assumptions: (1) the detector can be accurately calibrated by the legitimate user and remains isolated from the adversary, and (2) the detector noise is truly random from the adversary’s perspective. While significant research has focused on validating the first assumption—through improved calibration techniques and countermeasures against detector manipulation—little attention has been paid to validating the second. In fact, in QRNG community, it is commonly assumed that the electrical noise of an optical homodyne detector is not truly random and must be ``hashed out'' to extract genuine quantum randomness.

In this paper, we introduce a third option—the calibrated detector noise model—which lies between the untrusted and trusted noise models. This model relies solely on the first assumption: that the detector is isolated from Eve’s control. It allows for the possibility that detector noise is fully predictable to Eve. A key advantage is that it eliminates the need to validate the true randomness of detector noise, a task that can be challenging in practice. The trade-off is a slight reduction in QKD performance compared to the trusted noise model. Nevertheless, our numerical simulations show that the resulting loss in secret key rate is minimal under practical parameters. Specifically, when the detector noise variance is an order of magnitude below the vacuum noise ($\nu_B=0.1$), the proposed model achieves a secret key rate comparable to that of the trusted detector noise model. We also simulate the performance of a characterized commercial balanced detector ($\nu_B=0.28$), and show that even at a relatively high detector noise level, the calibrated model can still tolerate up to 20 dB of channel loss. Moreover, the new model can be readily incorporated into existing CV-QKD experimental setups.

\textbf{Acknowledgments} 
We thank an anonymous reviewer for pointing out a mistake in the autocorrelation calculation in an earlier version of the manuscript. We thank Shlok Nahar for comments on the trusted and untrusted detector noise models. Bing Qi acknowledges support from NYU-Shanghai start-up funds. 

\bibliographystyle{unsrt}

\bibliography{main}

\end{document}